\documentclass[aps,twocolumn,amssymb,amsfonts,amsmath,showpacs,final,a4paper,superscriptaddress]{revtex4-2}

\usepackage{float}
\usepackage[dvips,final]{graphicx}
\usepackage{dcolumn}
\usepackage{bm}
\usepackage[latin1]{inputenc}
\usepackage{textcomp}
\usepackage{amsmath}
\usepackage{color}

\begin{document}

\title{The ambivalent competition of Coulomb and van-der-Waals interactions in Xe--Cs$^+$ aggregates on Cu(111) surfaces}

\author{J. Thomas}
\affiliation{Fakult\"{a}t f\"{u}r Physik and Center for Nanointegration (CENIDE), Universit\"{a}t Duisburg-Essen, Lotharstr.~1, 47057 Duisburg, Germany}

\author{C. Bertram}
\affiliation{Fakult\"{a}t f\"{u}r Physik and Center for Nanointegration (CENIDE), Universit\"{a}t Duisburg-Essen, Lotharstr.~1, 47057 Duisburg, Germany}
\affiliation{Lehrstuhl f\"{u}r Physikalische Chemie\,I, Ruhr-Universit\"{a}t Bochum, Universit\"{a}tsstr.\,150, 44801~Bochum, Germany}

\author{J. Daru}
\affiliation{Lehrstuhl f\"{u}r Theoretische Chemie, Ruhr-Universit\"{a}t Bochum, Universit\"{a}tsstr.\,150, 44801~Bochum, Germany}

\author{J. Patwari}
\affiliation{Fakult\"{a}t f\"{u}r Physik and Center for Nanointegration (CENIDE), Universit\"{a}t Duisburg-Essen, Lotharstr.~1, 47057 Duisburg, Germany}
\affiliation{Lehrstuhl f\"{u}r Physikalische Chemie\,I, Ruhr-Universit\"{a}t Bochum, Universit\"{a}tsstr.\,150, 44801~Bochum, Germany}

\author{I. Langguth}
\affiliation{Lehrstuhl f\"{u}r Physikalische Chemie\,I, Ruhr-Universit\"{a}t Bochum, Universit\"{a}tsstr.\,150, 44801~Bochum, Germany}

\author{P. Zhou}
\affiliation{Fakult\"{a}t f\"{u}r Physik and Center for Nanointegration (CENIDE), Universit\"{a}t Duisburg-Essen, Lotharstr.~1, 47057 Duisburg, Germany}

\author{D. Marx}
\affiliation{Lehrstuhl f\"{u}r Theoretische Chemie, Ruhr-Universit\"{a}t Bochum, Universit\"{a}tsstr.\,150, 44801~Bochum, Germany}

\author{K. Morgenstern}
\affiliation{Lehrstuhl f\"{u}r Physikalische Chemie\,I, Ruhr-Universit\"{a}t Bochum, Universit\"{a}tsstr.\,150, 44801~Bochum, Germany}

\author{U. Bovensiepen}
\affiliation{Fakult\"{a}t f\"{u}r Physik and Center for Nanointegration (CENIDE), Universit\"{a}t Duisburg-Essen, Lotharstr.~1, 47057 Duisburg, Germany}

\email{uwe.bovensiepen@uni-due.de}

\date{\today}

\begin{abstract}
Microscopic insight into interactions is a key for understanding the properties of heterogenous interfaces. We analyze local attraction in non-covalently bonded Xe--Cs$^+$ aggregates and monolayers on Cu(111) as well as repulsion upon electron transfer. Using two-photon photoemission spectroscopy, scanning tunneling microscopy, and coupled cluster calculations combined with an image-charge model we explain the intricate impact Xe has on Cs$^+$/Cu(111). We find that attraction between Cs$^+$ and Xe counterbalances the screened Coulomb repulsion between Cs$^+$ ions on Cu(111). Furthermore, we observe that the Cs $6s$ electron is repelled from Cu(111) due to xenon's electron density. Together, this yields a dual, i.e., attractive or repulsive, response of Xe depending on the positive or negative charge of the respective counterparticle, which emphasizes the importance of the Coulomb interaction in these systems.
\end{abstract}

\maketitle
Properties of heterogenous interfaces are determined by covalent, Coulomb, and van-der-Waals interactions at an extent which is specific to the particular interface system. Competition or cooperation are decisive as is highlighted, for example, by a doping-induced, reentrant metal-insulator-metal phase transition in alkali doped K$_x$C$_{60}$ adsorbed on a Au(111) surface \cite{wachowiak2005,wang2007,feng2008}. While the individual fullerenes are interacting amongst each other and with the substrate by van-der-Waals interaction, the electrons donated by the alkalis determine the interface's electronic properties by exchange and Coulomb interaction \cite{feng2008,cui2019}. A similarly promising, yet unexplored pathway might be to prepare aggregates consisting of atoms as building blocks which allow a modification of the effective overall interaction by a variation of the atomic interactions. In this letter, we lay the foundations of this approach. We demonstrate a modification of the local interaction by coadsorbing Xe to Cs/Cu(111).

On metal surfaces the adsorption of noble gases like Xe is governed by an interplay of van-der-Waals attraction and Pauli repulsion \cite{bagus2002,dasilva2003}. Alkali atoms like Cs transfer, upon adsorption, their single valence electron to the metal surface \cite{langmuir1932,scheffler1991,gauyacq2007,politano2013}, being, in a good approximation, cations Cs$^+$ dominated by Coulomb interaction. The resulting surface dipole lowers the work function \cite{riffe1990,fischer1994,lu1996}. The investigation of coadsorbed Cs and Xe promises novel insights regarding competition (or cooperation) of these interactions. Such understanding is highly relevant from a fundamental viewpoint because local Coulomb interactions are essential in understanding structural order, electronic properties, and the elementary processes at heterogeneous interfaces. Those local effects and a potential control of the interactions are interesting for sensor design and energy conversion \cite{xu2015,schloegl2015}, amongst other applications. So far, electronic and ionic interactions in multi-component adsorbate systems are unknown on the single-particle scale.

Few percents of a monolayer (ML) Cs$^+$ form on Cu(111) a hexagonal alkali lattice due to mutual Coulomb repulsion of adjacent alkali ions observed in scanning tunneling microscopy (STM) \cite{vHofe2006}. The missing valence electron of the adsorbed alkalis induces an unoccupied electronic state observed in inverse photoemission \cite{jacob1987} and two-photon photoemission (2PPE) \cite{fischer1994,bauer1997}. The antibonding character of this Cs $6s$ state manifests  itself in an ultrafast energy shift due to the propagation of a nuclear wavepacket along the Cu--Cs$^+$ coordinate \cite{petek2000}.

Adsorption of noble gas layers on metal surfaces increases the lifetime of electrons in image potential states (IPS) in front of the surface \cite{mcneill1996,wolf1996} since their dielectric reponse repels the electronic wavefunction reducing the wavefunction overlap with electronic states of the substrate \cite{hotzel1999,bertholdt2004}. Though well-established in two dimensions, this phenomenon remains to be explored for structures of lower dimensionality. This holds even more since the highly polarizable Xe is considered as a non-polar solvent in the presence of cations \cite{rentzepis1981}. The fundamental question of interest here is how Xe affects the structural and electronic properties of alkali ions on surfaces.

\begin{figure}[htb]
\includegraphics[width=0.99\columnwidth]{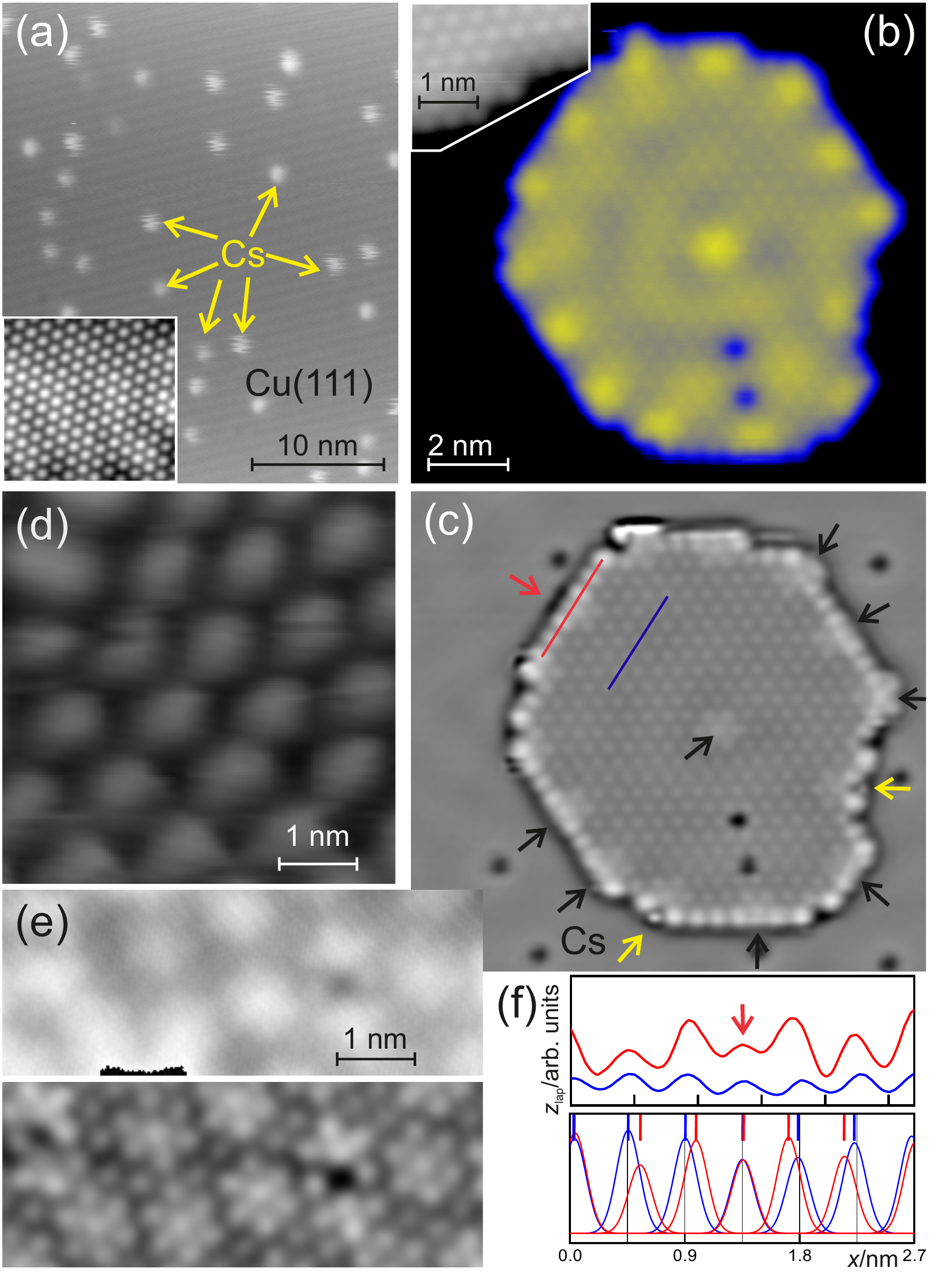}
\caption{STM images (a) of Cs$^+$/Cu(111), -500~mV, 10~pA, and (b) of a Xe-Cs$^+$ aggregate on Cu(111) using a false-color scale, 10~pA, -25~mV. Insets in (a,b) show 1~ML Xe/Cu(111), recorded at 1.2~nA, 100~mV on a wetting ML and at an aggregate perimeter, 44~pA, 7 mV, respectively. (c) Laplace-filtered image of (b). The Xe:Cs$^+$ ratio in (b,c) is 40:1. (d) 0.16~ML Cs$^+$/Cu(111), 10~pA, 250~mV; (e) Original (top) and Laplace-filtered image (bottom) of 1~ML Xe on top of 0.16~ML Cs$^+$/Cu(111), 89~pA, 250~mV, (f) apparent height profiles (top) and their Gaussian fits with maxima marked (bottom) in panel (c) in identical colors.
\label{fig_1}}
\end{figure}

In this letter, we unveil that Xe mediates an effective, attractive interaction between Cs$^+$ on the surface \emph{and} enhances attraction between Cs$^+$ and Cu(111). Furthermore, Xe induces a repulsion of the Cs $6s$ electron from Cu(111). We trace this dual impact of Xe back to its polarization of opposite sign induced by cations and electrons. This was achieved by using STM and 2PPE as complementary tools which probe Cs$^+$ and Cs, respectively, on a wetting Xe ML with Cs$^+$ on Cu(111), supplemented by aggregates at sub-ML Xe. Supported by coupled cluster calculations in conjunction with image-charge modeling of the screened Coulomb repulsion, we showcase the impact of the dominant Coulomb interaction locally.

The experiments were performed in two separate ultrahigh vacuum systems at a base pressure of $4\cdot 10^{-10}$~mbar. STM was carried out at $T\leq8$~K. 2PPE spectra were measured at $T=30$~K using two time-delayed laser pulses of 3.1~eV photon energy, 40~fs pulse duration at an incident fluence of $1$~mJ/cm$^2$ \cite{sandhofer2014}. The Cu(111) surfaces were prepared by sputter-anneal cycles. The Cs atoms were deposited at $T=200$~K from commercial Cs dispensers (SAES getters). The alkali coverage was determined by counting the ions on Cu(111) using STM and an analysis of the work function in 2PPE \cite{fischer1994,lu1996}. Xenon was deposited from a background pressure of $2\cdot 10^{-6}$~mbar in the STM and exposed to a Xe beam in the 2PPE setup. The Xe coverage was determined from temperature programmed desorption calibrated by desorbing the first ML from Cu(111). All electronic structure calculations have been performed using the ORCA package. The energies were computed using canonical CCSD(T) theory employing complete basis set (CBS) extrapolation \cite{supplement}.

After Cs deposition, protrusion cover the surface randomly at an apparent coverage of $7\cdot 10^{-4}$~ML, which are identified as Cs$^+$, Fig.~\ref{fig_1}(a). Cs$^+$ is partially pinned at defects and partially mobile, leading to a striped appearance. After Xe adsorption Cs$^+$ ions are trapped in Xe-Cs$^+$ aggregates, Fig.~\ref{fig_1}(b,c). The aggregate's flat parts exhibit the same R30$^\circ-\sqrt{3}\times \sqrt{3}$ hexagonal superstructure as pure Xe/Cu(111), see insets in Fig.~\ref{fig_1}(a,b). We thus assign the flat parts of the aggregates to single Xe layers. The aggregates exhibit protrusions at the border and in the center, that do not exist for pure Xe layers, Fig.~\ref{fig_1}(b) inset. The Laplace filtered image in panel (c) identifies two causes. For some, marked by yellow arrows in Fig.~\ref{fig_1}(c), a small feature with the lateral size of a Cs$^+$ is surrounded laterally by Xe atoms, which are imaged at a larger apparent height than other Xe atoms. For other protrusions, marked by black arrows in (c), the lateral size of all features is similar. We assign these latter protrusions to Cs$^+$ covered by Xe. Both cases represent 2D solvation of Cs$^+$, but of different coordination. In the aggregate center, one Cs$^+$ is fully solvated, showing up in the Laplace-filtered image as a Xe heptamer. Such heptamers are the dominant species in STM images of 1~ML~Xe/0.16~ML~Cs$^+$/Cu(111) in Fig.~\ref{fig_1}(e). Their number reflects the Cs$^+$ coverage. The uniformity of the solvatomers facilitates an analysis by a space averaging method like 2PPE reported below. We explain the observed structure by a carpet-like growth \cite{foelsch1989,matthaei2012} of the Xe layer across Cs$^+$ which is facilitated by the larger size of Xe  ($r_{\mathrm{Xe}} = 216$~pm vs.\ $r_{\mathrm{Cs^+}}=174$~pm) and its larger physisorption distance of 360~pm \cite{seyller1998} compared to the chemisorption distance of Cs$^+$ of 301~pm \cite{lindgren1983}.

Height profiles across a row of Xe atoms in the interior of the aggregate reflect the regular distance between two Xe of 0.45~nm, blue line in  Fig.~\ref{fig_1}(f). The distances between the maxima in the height profile across a Xe-covered Cs$^+$ are less regular and indicate a nearest neighbor Cs$^+$--Xe distance which is shorter by $\sim 50$~pm, red line and vertical markers ibid, better observed in the Gaussian fits to the height profiles (bottom). This leads to a larger distance to the next nearest neighbor to maintain registry with the Xe environment \cite{note_rim}.

\begin{figure}[htb]
\includegraphics[width=0.99\columnwidth]{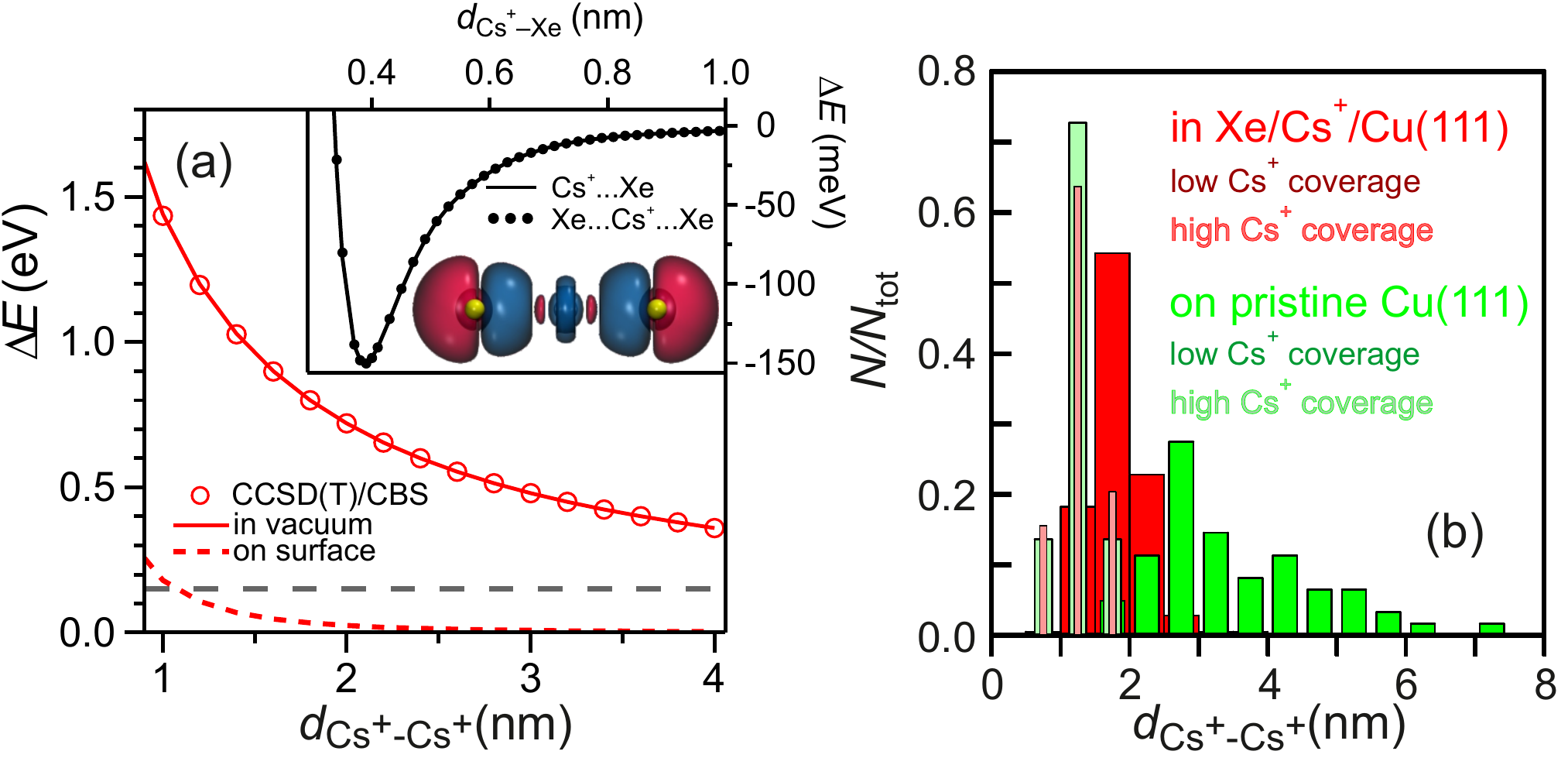}
\caption{Cs$^+-$Cs$^+$ distances in Xe$-$Cs$^+$ aggregates: (a) Repulsive interactions between two Cs$^+$ without (solid line) and with contact to a metal (red dashed line) as a function of the \mbox{Cs$^+-$Cs$^+$} distance. Circles represent the CCSD(T)/CBS reference. Inset: Attractive interactions between Cs$^+$ and one/two Xe~atoms per Xe obtained from CCSD(T)/CBS calculations as a function of \mbox{Cs$^+-$Xe} (equi)distance; the resulting attraction of 150~meV per nearest-neighbor Xe~atom provides the gray dashed line in the main panel. The blue/red isosurfaces of the charge density difference visualize charge accumulation/depletion of \mbox{Xe$-$Cs$^+-$Xe} at the optimized equidistance of 0.39~nm. Xe cores are yellow \cite{supplement}. (b) Nearest-neighbor distance histogram of Cs$^+$ on Cu(111) in green and in Xe$-$Cs$^+$ aggregates in red normalized to the number of distances $N_{\mathrm{tot}}$ in STM images for $\Theta_{\mathrm{Cs^+}}=7\cdot 10^{-4}$~ML (low) and $0.16$~ML (high).
\label{fig_2}}
\end{figure}

The formation of such aggregates is surprising as aggregates containing multiple positive ions are expected to break up into smaller ones due to Coulomb repulsion. We shed light on this paradox by CCSD(T)/CBS calculations to compare the Cs$^+-$Xe attraction to the Cs$^+-$Cs$^+$ repulsion in the absence and presence of screening by a metal surface, see Fig.~\ref{fig_2}(a). At the preferred Cs$^+-$Xe distance of 0.4~nm, which is in accord with the measured distances, see Fig.~\ref{fig_1}(f), the attractive interactions between Cs$^+$ and Xe amounts to $\sim150$~meV in vacuum, see Fig.~\ref{fig_2}(a), inset. We conclude that although this Cs$^+-$Xe attraction features significant electronic polarization as demonstrated by our charge density difference analysis, Fig.~\ref{fig_2}(a) inset, in addition to dispersion, it is one order of magnitude smaller than the unscreened Coulomb repulsion between two Cs$^+$ at observed separations of $d_{\rm Cs^+- Cs^+} \approx 1$--2~nm in the aggregates, see Fig.~\ref{fig_2}(b). Thus, without a metal support, Xe$-$Cs$^+$ aggregates should disperse into smaller ones in contrast to our observations.

To scrutinize the role of Cu(111) we have factored in image charges \cite{ic1,ic2}. Having shown that the Coulomb repulsion between point charges represents the Cs$^+-$Cs$^+$ interactions as given by explicit electronic structure calculations faithfully at the experimentally relevant distances (compare solid line to circles in Fig.~\ref{fig_2}(a)), we discuss the impact of image charges at the metal surface on equal footing \cite{supplement}. As evidenced by Fig.~\ref{fig_2}(a) the screening effect is enormous at 1.5~nm: The Coulomb repulsion is decreased from $\approx 1$~eV (solid line) to $\approx 50$~meV (red dashed line), which is much smaller than the Cs$^+-$Xe attraction of 150~meV per Xe~atom (gray dashed line).

The distances between two Cs$^+$ have been analyzed for low and high Cs$^+$ coverage $\Theta_{\mathrm{Cs^+}}$ in the absence (presence) of Xe yielding the green (red) data in Fig.~\ref{fig_2}(b). For low $\Theta_{\mathrm{Cs^+}}$ a sharp distribution of Cs$^+-$Cs$^+$ pairs at short distances of about 1--2~nm is observed within Xe-Cs$^+$ aggregates for different Xe coverages \cite{supplement}. In this limit the screened Coulomb repulsion between two Cs$^{+}$ on Cu(111) is overcompensated by the attractive interaction between Cs$^+$ and Xe, as revealed by the calculations. Phenomenologically an effective attraction between cations on Xe/Cu(111) with reference to the pristine Cu(111) surface is obtained. The picture extracted is that individual Cs$^{+}$/Cu(111) prefer to reside at Xe aggregates given their favorable interactions with Xe. Being bound to the rim, the remaining weak repulsion between any two such cations leads to similar Cs$^+-$Cs$^+$ distances along the perimeter.

\begin{figure}
\includegraphics[width=0.99\columnwidth]{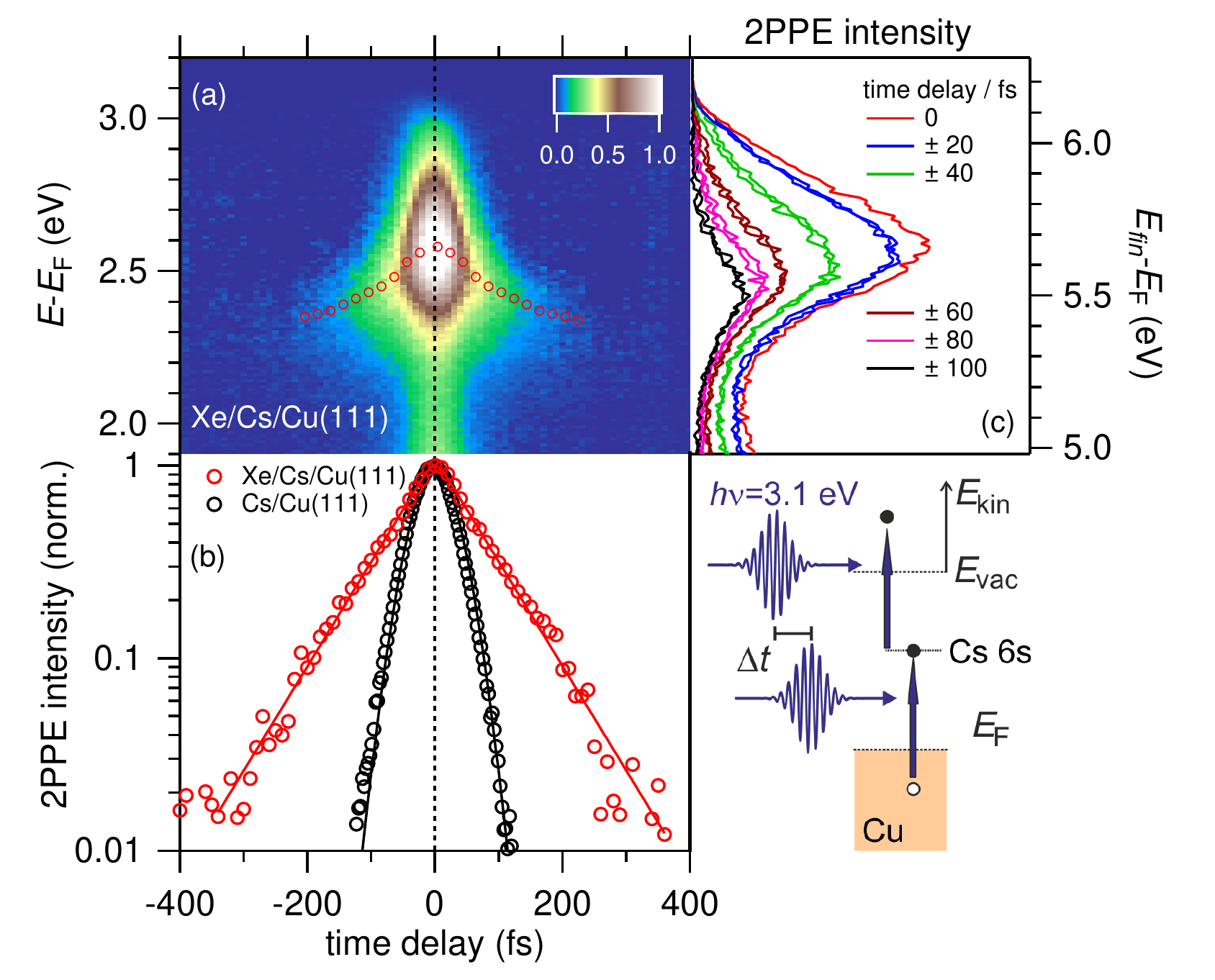}
\caption{2PPE intensity (a) of 1~ML Xe/0.16~ML Cs$^+$/Cu(111) recorded at $T=30$~K in false color as a function of intermediate energy $E-E_{\mathrm{F}}$ (left), final state energy (right), and time delay. The time-dependent energy of the Cs $6s$ feature is indicated by ($\circ$). (b) Autocorrelation traces of the Cs $6s$ state ($\circ$), obtained by integrating the intensity within 200 and 150~meV for Cs/Cu(111) (black) and Xe/Cs/Cu(111) (red), respectively. Relaxation times of $(13\pm3)$~fs and $(80\pm10)$~fs for Cs/Cu(111) and Xe/Cs/Cu(111), respectively, are obtained by fits of single exponential decays convolved with the autocorrelation of the laser pulse (---) \cite{note_tau}. (c) 2PPE spectra at the indicated time delays.
\label{fig_3}}
\end{figure}

At higher $\Theta_{\mathrm{Cs^+}}$ a regular superstructure forms with an average Cs$^+$--Cs$^+$ distance of $(1.25 \pm 0.22)$~nm, which is smaller than the distance reported for the aggregates, Figs.~\ref{fig_1}(d), \ref{fig_2}(b). Xe adsorption leads to a regular array of heptamers, also found in the aggregate center in Fig.~\ref{fig_1}(c), with the same distance of $(1.28 \pm 0.28)$~nm, Figs.~\ref{fig_1}(e), ~\ref{fig_2}(b). As for low $\Theta_{\mathrm{Cs^+}}$, the Xe overgrows the Cs$^+$ like a carpet, Fig.~\ref{fig_1}(e). Within the carpet the Xe--Cs$^+$ interaction also modifies the adjacent Xe--Xe distance \cite{note_rim}. These are the surfaces that were investigated by time-resolved 2PPE to provide insight in the response of the environment to this local excitation \cite{gauyacq2007,zhao2008}. As sketched in Fig.~\ref{fig_3}, the photon energy is set for resonant electron attachment from Cu(111) to the Cs $6s$ state \cite{bauer1997,petek2000}. Adsorption of 1~ML Xe increases the work function of 0.14~ML Cs$^+$/Cu(111) by $(100\pm15)$~meV. The energy of the Cs $6s$ feature varies by $(120\pm15)$~meV \cite{supplement}. This behavior suggests a Xe induced decoupling of the Cs $6s$ electron from Cu(111), contrary to the effective attraction Xe mediates on the metal surface between two Cs$^+$ reported above.

\begin{figure}
\includegraphics[width=0.99\columnwidth]{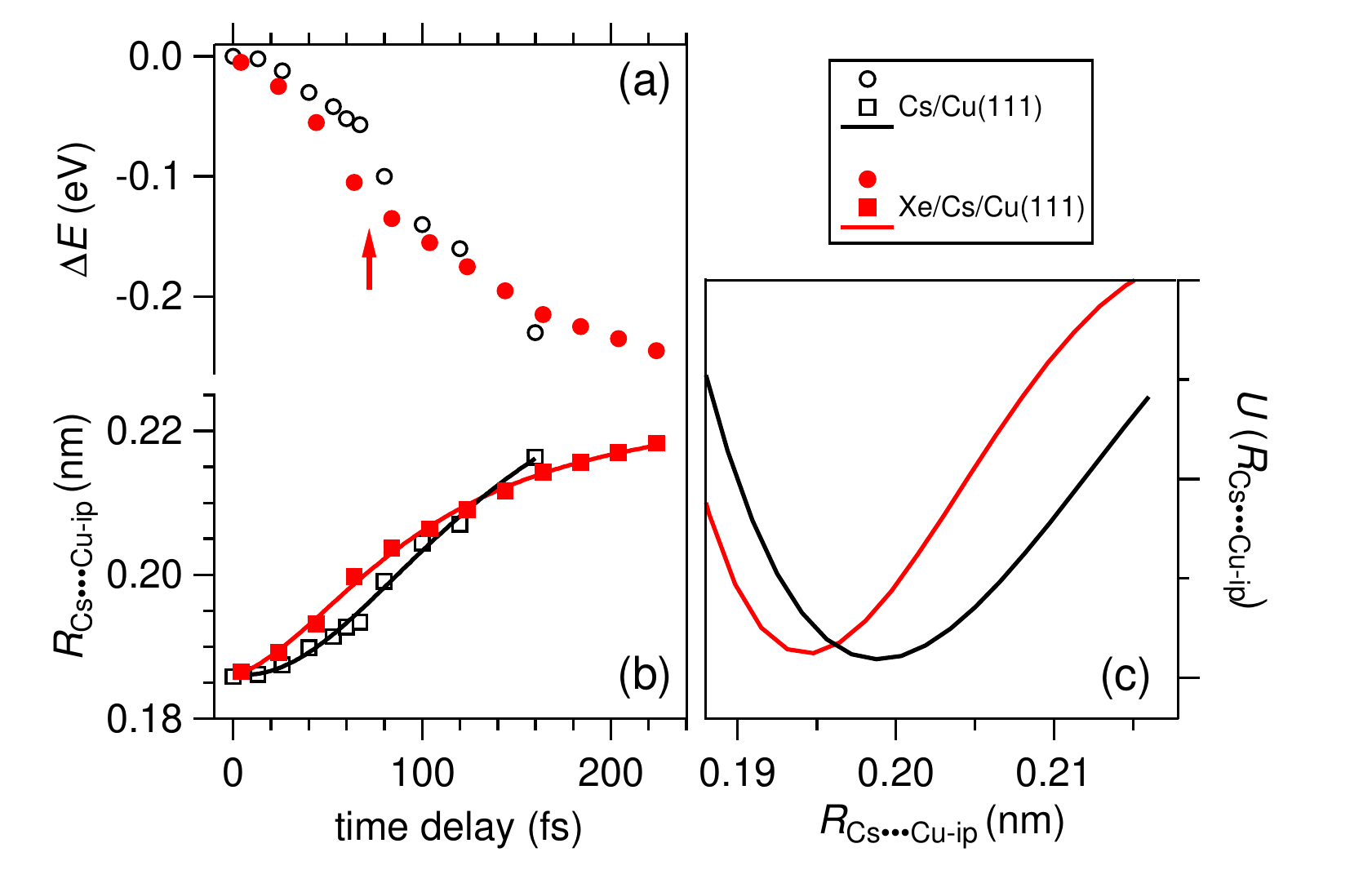}
\caption{Time-dependent change of intermediate state energy $\Delta E$ (a) plotted for Xe/Cs$^+$/Cu(111) from Fig.~\ref{fig_3} and for bare Cs$^+$/Cu(111) using data from \cite{petek2000}. The respective distances $R_{\mathrm{Cs\cdots Cu-ip}}$ and the excited state potential energy surface $U(R_{\mathrm{Cs\cdots Cu-ip}})$ are determined with (red) and without (black) coadsorbed Xe. Note that the absolute energy difference between the two potentials in (c) is not know.
\label{fig_4}}
\end{figure}

Time-resolved 2PPE provides microscopic insights into this decoupling. Fig.~\ref{fig_3}(a) shows the 2PPE intensity autocorrelation as a function of time delay $\Delta t$ between two identical laser pulses. The Cs $6s$ feature occurs at time zero at $E-E_{\mathrm{F}}=2.6$~eV and relaxes to lower energy with increasing $\Delta t$ due to the progression of the nuclear wave packet on the anti-bonding potential energy surface \cite{petek2000}. Two effects of Xe are identified. First, the Cs $6s$ electron relaxation time $\tau$ is increased by a factor of 6 from 13 to 80 fs, Fig.~\ref{fig_3}(b) and \cite{note_tau}. Since such a pronounced increase in $\tau$ represents a substantial reduction in the overlap of the Cs $6s$ wavefunction with Cu(111) states \cite{gauyacq2007}, we conclude that Xe coadsorption induces a decoupling of the Cs $6s$ wavefunction from Cu(111). Second, while the observed transient energy gain is 0.24~eV, similar to Cs/Cu(111) \cite{petek2000}, Xe influences the time dependent energy changes {$\Delta E$}, Fig.~\ref{fig_4}(a). An inflection point identified for Xe/Cs/Cu(111) at $\Delta t=70$~fs, see arrow, is absent for Cs/Cu(111). Following \cite{petek2000} we estimate the potential along the surface normal direction from $\Delta E(\Delta t)$. We convert the energy to the distance $R_{\mathrm{Cs\cdots Cu-ip}}$ between Cs and the image-plane of the Cu(111) interface using previous calculations \cite{borisov1999}. We fit the values by an analytical function, Fig.~\ref{fig_4}(b), under the assumption of the same $R_{\mathrm{Cs\cdots Cu-ip}}(E)$ because Xe has a minor effect on the Cs 6$s$ energy. Using $\partial R^2/ \partial \Delta t^2$, which defines the force $F$ acting on the Cs ion core, we determine the excited state potential $U(R_{\mathrm{Cs\cdots Cu-ip}})=-\int{F \cdot dR_{\mathrm{Cs\cdots Cu-ip}}}$. The minimum in $U$ shifts $(4.0\pm1.5)$~pm towards Cu(111) upon adding Xe, resembling the attraction Xe mediates among two Cs$^+$ concluded above, Fig.~\ref{fig_4}(c).

Our analysis establishes that adding Xe to Cs$^+$/Cu(111) compensates for the repulsive interaction in the surface plane between adjacent Cs$^+$ and enhances the binding of Cs$^+$ to Cu(111) along the surface normal. The observed increase in Cs $6s$ electron lifetime upon adding Xe is explained by a decrease in wavefunction overlap with the states of Cu(111), which implies that the respective Cs wavefunction $\Psi_{6s}$ is repelled by Xe from Cu(111) to the vacuum. Such repulsion has been concluded previously from an increase in electronic lifetimes of IPS upon adsorbing a closed noble gas layer on metal surfaces \cite{mcneill1996,wolf1996,hotzel1999,bertholdt2004}. It is based on the dielectric polarization response of the noble gas electron density to the transient IPS population. In the present case, Cs$^+$ is surrounded by Xe (Fig.~\ref{fig_1}). Since $\Psi_{6s}$ extends laterally over several Cu lattice constants \cite{gauyacq2007} and affects the appearance of
adjacent Xe atoms in STM images (Fig.~\ref{fig_1}), we explain the repulsion of $\Psi_{6s}$ by the part of the Xe electron density that is beneath Cs$^+$ which reduces the overlap of $\Psi_{6s}$ with Cu(111) states. While on homogenous surfaces in two spatial dimensions 1~ML of Xe doubles the IPS lifetime \cite{bertholdt2004}, the larger, sixfold increase in $\tau$, see Fig.~\ref{fig_3}(b), is explained by the localized nature of $\Psi_{6s}$. The lifetime increase for Xe/Cs/Cu(111) is attributed to the reduced dimension where seven Xe atoms
modify a single Cs$^+$, while for the two-dimensional layers the homogenous electron density acts and provides less electron density for screening.

In conclusion, we have investigated the mutual influence of Xe and Cs$^+$ on a Cu(111) surface and discovered two-dimensional, nanostructured Xe-Cs$^+$ aggregates. Weak interactions with Xe outperform the Coulomb repulsion between cations due to screening by the metallic surface. A proposed electron repulsion between the Xe electron density and the Cs $6s$ wave function corroborates the dominant character of the local Coulomb interaction. We believe that our results are of general nature. Since alkali and noble gases represent groups in the periodic table, systematic variation of the described local interactions could become feasible by changing the atomic polarizability via the noble gas and the different surface dipoles via alkalis. Extension to earth alkalis and halogens bears additional potential. Since the interfacial electronic structure and the constituent's electron affinity contribute as well \cite{bertholdt2004} further means for modification exist. The atomic-scale understanding established here provides opportunities to tune local, non-covalent interactions and opens the doorway toward establishing novel nanostructuring strategies of heterogenous interfaces.

\begin{acknowledgments}
Funded by the Deutsche Forschungsgemeinschaft (DFG, German Research Foundation) under Germany's Excellence Strategy - EXC 2033 - 390677874 - RESOLV and by Project-ID 278162697 - SFB 1242.
\end{acknowledgments}


\begin{thebibliography}{99}
\bibitem{note_rim}In the 1~ML Xe/0.16~ML~Cs$^+$case this Cs$^+$-induced shortening of the Xe-Xe distanc is particularly visible for adjacent heptamers. The shortened distance within the heptamers leads to a larger distance between the Xe of adjacent heptamers, visible as a dark rim around them, see Fig. 1(e).
\end{thebibliography}
\end{document}